\documentclass{an}
\usepackage{graphicx}
\usepackage{times}
\usepackage{fancyhdr}
\begin{document}

\title{AB Doradus C: Age, Spectral Type, Orbit, and Comparison to Evolutionary Models}

\author{E.L. Nielsen\inst{1}\fnmsep\thanks{E. Nielsen acknowledges the 
support of the Michelson Science Center Graduate Student Fellowship.} 
\and  L.M. Close\inst{1}
\and  J.C. Guirado\inst{2}
\and  B.A. Biller\inst{1}
\and  R. Lenzen\inst{3}
\and  W. Brandner\inst{3}
\and  M. Hartung\inst{4}
\and  C. Lidman\inst{4}}
\institute{
Steward Observatory, University of Arizona, 933 N. Cherry Ave, Tucson, AZ 85721 USA
\and Departament d'Astronomia i Astrofisica, Universitat de Valencia, E-46100 Burjassot, Valencia, Spain
\and Max-Plank-Institut f\"{u}r Astronomie, K\"{o}nigstuhl 17, D-69117 Heidelberg, Germany
\and European Southern Observatory, Alonso de Cordova 3107, Santiago 19, Chile}

\date{Received $<$date$>$; 
accepted $<$date$>$;
published online $<$date$>$}

\abstract{We expand upon the results of Close et al. 2005 regarding the 
young, low-mass object AB Dor C and its role as a calibration point for 
theoretical tracks.  We present an improved spectral reduction and a new 
orbital solution with two additional epochs.  Our improved analysis confirms 
our spectral type of M8 ($\pm$1) and mass of 
0.090 $\pm$0.003 M$_{\sun}$ for AB Dor C.  Comparing the results for AB Dor 
C with other young, low-mass objects with dynamical masses we find a 
general trend where current evolutionary models tend to over-predict 
the temperature (or under-predict the mass) for low mass stars and brown 
dwarfs.  Given our precision, there is a $\sim$99\% chance that the mass of 
AB Dor C is underestimated by the DUSTY tracks in the HR diagram.
\keywords{Infrared: Stars, Stars: Formation, Stars: Low-Mass, Brown Dwarfs, Stars: Pre-Main-Sequence, Stars: Individual: AB Dor}
}

\correspondence{enielsen@as.arizona.edu}

\maketitle

\section{Introduction}

The study of young, low-mass objects has been yielding increasingly fruitful 
science, yet the field remains dependent on evolutionary models to 
properly interpret the data that are collected from these objects.  In 
particular, mass, while a fundamental property, is very rarely measured 
directly, and instead must be inferred from theoretical tracks (e.g., 
Burrows, Sudarsky, and Lunine 2003, Chabrier et al. 2000).  It is thus of 
great interest to find calibrating objects that can link a dynamically 
measured mass with observables such as NIR (1-2 $\mu$m) fluxes and 
spectral types.

In our previous work (Close et al. 2005), we reported the direct detection 
of the 
low-mass companion to the young star AB Dor A, along with measurements of 
the JHKs fluxes, spectral type, and dynamically determined mass of 
AB Dor C.  Upon 
comparing these results with the predictions of Chabrier et al. 2000, we 
found the models to be systematically over-predicting the fluxes and 
temperature of AB Dor C, given an age of the system of 50 Myr.  Put 
another way, the model masses seem 
to be underestimating the mass of a low-mass 
object given its age, NIR fluxes, and spectral type.  Since the publication 
of these results, another calibrating object has been reported by 
Reiners, Basri, and Mohanty 2005: USco CTIO 5.  While this equal-mass binary 
is younger ($\sim$8 Myr) and more massive (total mass $\ge$0.64 M$_{\sun}$) 
than AB Dor C, 
Reiners et al. find the same trend of models under-predicting masses based 
simply on photometric and spectroscopic data applied to the HR diagram.  
A similar trend for such 
masses was previously noted by Hillenbrand and White 2004.  Moreover, this 
trend has been theoretically predicted for higher masses by Mohanty, 
Jayawardhana, and Basri 2004, and by Marley et al. 2005 for planetary masses.

In this paper, we seek to expand on our earlier results from Close et al. 
2005, using an improved spectral reduction and a more robust determination 
of the spectral type.  We also present an improved orbital fit based on 
additional astrometric data, as well as address concerns raised by 
Luhman, Stauffer, and Mamajek 2005 regarding the age of the AB Dor system.

\section{An Improved Spectral Reduction}

As described in Close et al. 2005, in February 2004 we obtained 20 minutes 
of K-band spectra using the Very Large Telescope (VLT), following our initial 
detection of AB Dor C.  Using the R $\approx$ 1200-1500 (2-2.5 $\mu$m) grism 
and the 0.027'' pixel camera of NACO (see Lenzen et al. 2003), the 0.086'' 
slit was aligned along the centers of both AB Dor A and C.  The observations 
themselves consisted of eight deep exposures, intentionally saturating the 
inner 
pixels of the spectral PSF, with the two objects (A and C) nodded 
along the slit between exposures.  An additional eight exposures were 
obtained with a 180$^{\circ}$ rotation of the derotator, flipping the 
relative positions of A and C.

Given our measured separation of A and C of 0.156'' (5.78 pixels), and a flux 
ratio at Ks of 80, the signal from AB Dor C lies beneath the wings of the 
PSF of A.  Relying on the relative stability of the NACO PSF, we subtracted 
a 0$^{\circ}$ image from a 180$^{\circ}$, removing the signal from A 
while leaving a positive and negative spectrum of AB Dor C, which can then be 
extracted easily using the standard Image Reduction and Analysis Facility 
(IRAF) routines.  This aligning of the PSFs is complicated by sub-pixel 
variations in the order position and orientation from image to image, as 
well as by variations in the total flux across different exposures.

To prepare for this subtraction, the first step is to align the 0$^{\circ}$ 
and 180$^{\circ}$ images.  This pair of images 
is considered one dispersion pixel at a time (that is, we consider the 
1024 spatial pixels along the chip corresponding to a single wavelength 
element), where one image undergoes a series of sub-pixel shifts before the 
two one-dimensional segments are subtracted.  A series of reference pixels 
between 9 and 12 pixels (0.24'' and 0.32'') from the center of the 
PSF are chosen so as to avoid the saturated central 
pixels, and any signal from AB Dor 
C (which would be asymmetric from 0$^{\circ}$ to 180$^{\circ}$).  The 
minimum variance in these reference pixels after subtraction is taken to 
correspond to the best shift.  This process is repeated for each of the 1024 
dispersion pixels, a polynomial is fit to these individual shifts, and 
this fit is used to prepare the final subtraction.

\begin{figure}
\resizebox{\hsize}{!}
{\includegraphics[width=3in]{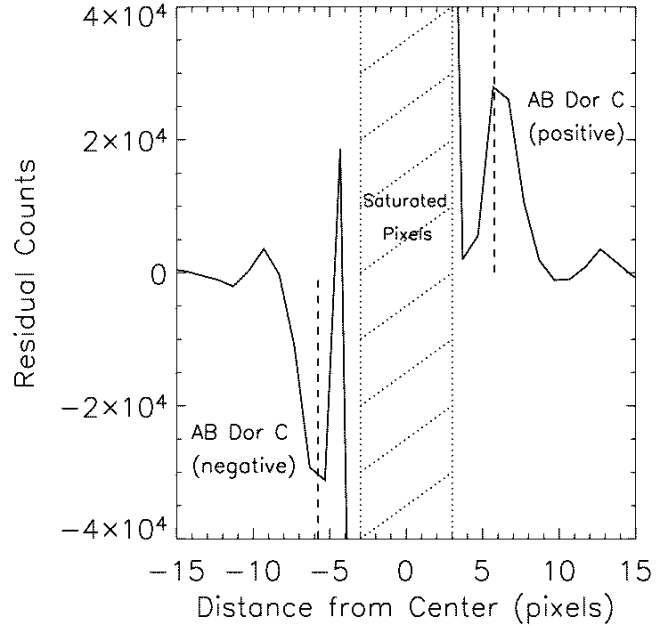}}
\caption{An example of a subtraction of a 0$^{\circ}$ image from a 
180$^{\circ}$, with each spatial pixel along the x-axis representing a sum 
along the dispersion direction.  The expected positions of AB Dor C on either 
side are marked with dashed lines; the clear positive and negative signals 
at this location indicate a good subtraction of AB Dor A.  
The spectrum is extracted 
from both the positive peak on the right, as well as the negative peak on 
the left.  The marked central region (where the data are not plotted) 
indicate the saturated pixels which were ignored during the reduction.}
\label{subfig}
\end{figure}

We continue this procedure iteratively with various values of a global flux 
scaling of one of the two images, again minimizing the variance in the 
reference pixels to find the best fit.  An example of this subtraction of a 
0$^{\circ}$ and 180$^{\circ}$ image is shown in Figure~\ref{subfig}.  Each 
point represents a sum across all 1024 dispersion pixels corresponding to 
a given spatial pixel, with the dashed lines indicating the location of AB 
Dor C.  The inner, saturated pixels show a large amount of noise as would 
be expected, but at the position of AB Dor C are obvious negative and 
positive peaks.  We repeat these steps for each possible pairing of 
the eight 0$^{\circ}$ and eight 180$^{\circ}$ spectra, eliminating pairs 
where the routine did not converge, and choosing the best subtraction; this 
yields a total of 13 spectra (out of 16 possible), which are then 
combined.

We note that while there are many pairs where this process 
fails to converge, we never see a ``false-positive'' signal.  That is, while 
for many of our spectra we see a positive peak to the right and a negative 
peak to the left, as seen in Figure~\ref{subfig}, not even once (out of 64 
trials) do we see a positive peak on the left or a negative peak on the 
right at the position of AB Dor A.  This leaves us confident that we have 
extracted the signal from AB Dor C, rather than spurious light.  
Additionally, these 13 independent extractions of the spectra are all 
qualitatively similar; each appears as a late-M spectrum, clearly 
distinguishable from the spectrum of AB Dor A.

Observations of a 
standard star (HIP 24153, G3V) taken within a half hour of the AB Dor images 
are used to remove telluric lines, and a modified solar spectrum compensates 
for stellar features from the standard (Maiolino, Rieke \& Rieke 1996).

\begin{figure}
\resizebox{\hsize}{!}
{\includegraphics[width=3in]{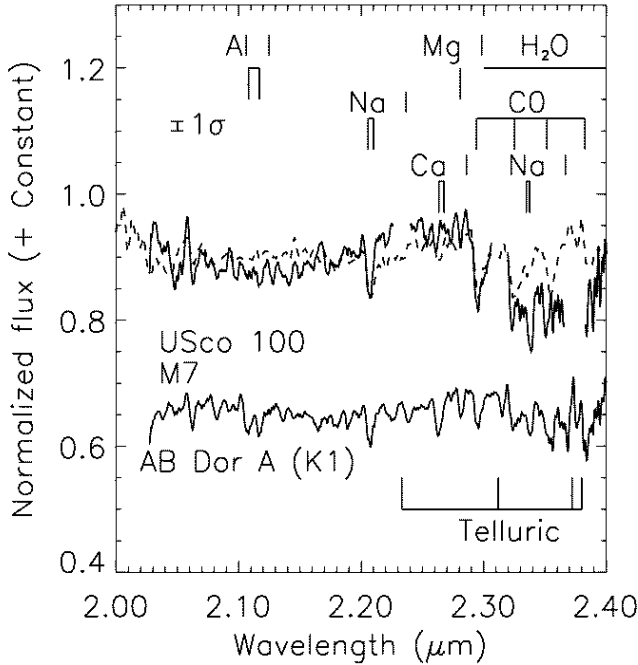}}
\caption{The spectrum of AB Dor C (upper solid line) shown against that 
of USco 100 (dashed line), a young ($\sim$8 Myr) 
M7 (Gorlova et al. 2003), with the continuum of both objects removed.  
Features arising from an incomplete removal of telluric lines 
are marked, and are not plotted in the AB Dor C spectrum.  The strength 
of C's sodium line at 2.21 $\mu$m and the depth of the H$_2$O absorption 
and the first 
CO break at 2.3 $\mu$m suggest AB Dor C is cooler than an M7, at the 
1$\sigma$ level of the observational noise, as indicated in the figure (this 
noise, 0.015, was found by taking the standard deviation of the AB Dor C 
spectrum between 2.13 and 2.18 $\mu$m, a featureless section of spectrum 
between the Al I and Na I doublets).  The 
spectrum of AB Dor A (which was used as a reference for poorly-removed 
telluric features) is also shown at the bottom of the plot.}
\label{spec01_fig}

\end{figure}

\begin{figure}
\resizebox{\hsize}{!}
{\includegraphics[width=3in]{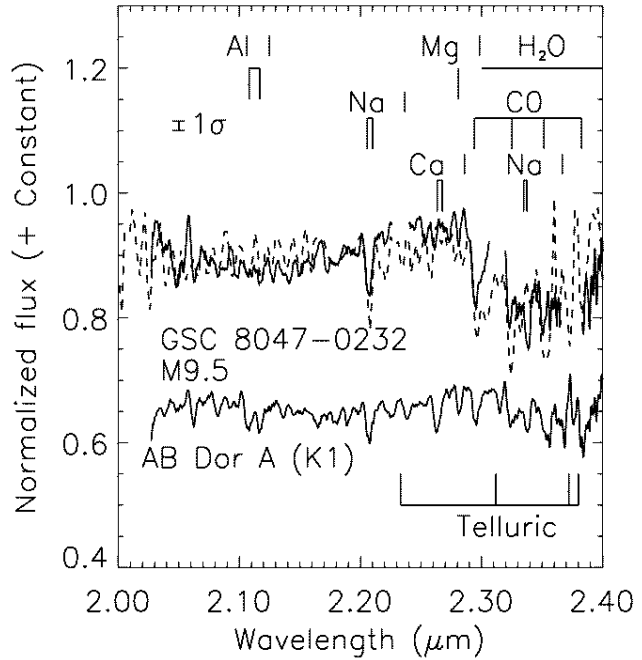}}
\caption{The spectrum of AB Dor C, this time plotted with GSC 
8047-0232, a young ($\sim$ 30 Myr) M9.5 
(Chauvin et al. 2005).  The sodium and 
CO features of the template now appear cooler than those of AB Dor C, 
bounding the spectral type between M7 and and M9.5, at the 1$\sigma$ 
level.}
\label{spec02_fig}
\end{figure}

The final spectrum is shown plotted against two young, late-M templates in 
Figures~\ref{spec01_fig} and~\ref{spec02_fig}.  Since we 
were unable to preserve the continuum 
of AB Dor C through our data reduction, we simply remove the continuum from 
our spectrum as well as that of the template (using polynomial fits of the 
same order).  Judging by the depth of the CO breaks, and the 
strength of the Na I line at 2.21 $\mu$m, these templates constrain the 
spectral type of AB Dor C between M7 and M9.5 at the 1$\sigma$ level, as 
was previously 
reported in Close et al. 2005.  Additionally, we found four features in the 
spectrum that do not seem consistent with any late-M.  We compared our 
unsaturated spectrum of AB Dor A to other K1 spectra, and finding these 
features present in A as well, we determined these lines to be telluric 
features that were not fully removed, hence we have not plotted these 
small segments of the spectra.


We note in Figures~\ref{spec01_fig} and~\ref{spec02_fig} that while both 
AB Dor A and C show a Na I line at 2.21 $\mu$m and CO features between 2.3 
and 2.4 $\mu$m, AB Dor A shows a strong 2.26 $\mu$m Ca I triplet absorption 
feature while AB Dor C does not.  Similarly, there is no correlation between 
the strength of the Mg I line (2.28 $\mu$m) or the Al I doublet 
(2.11 $\mu$m) between A and C, as would occur if our spectrum were dominated 
by spurious light from AB Dor A.  
Comparing the line strengths, we find the equivalent width 
of the Na I 2.21 $\mu$m doublet in C to be $\sim$1.5 times that of A.  
Meanwhile, 
the Ca I 2.26 $\mu$m feature equivalent width for C is less 
than 5\% of A.  This 
leaves us confident that the amount of contamination from AB Dor A is 
$\le$5\%, with similar results obtained from analysis of the Al I doublet.

\section{Improved Orbit}

\begin{figure*}
\resizebox{\hsize}{!}
{\includegraphics[width=3in]{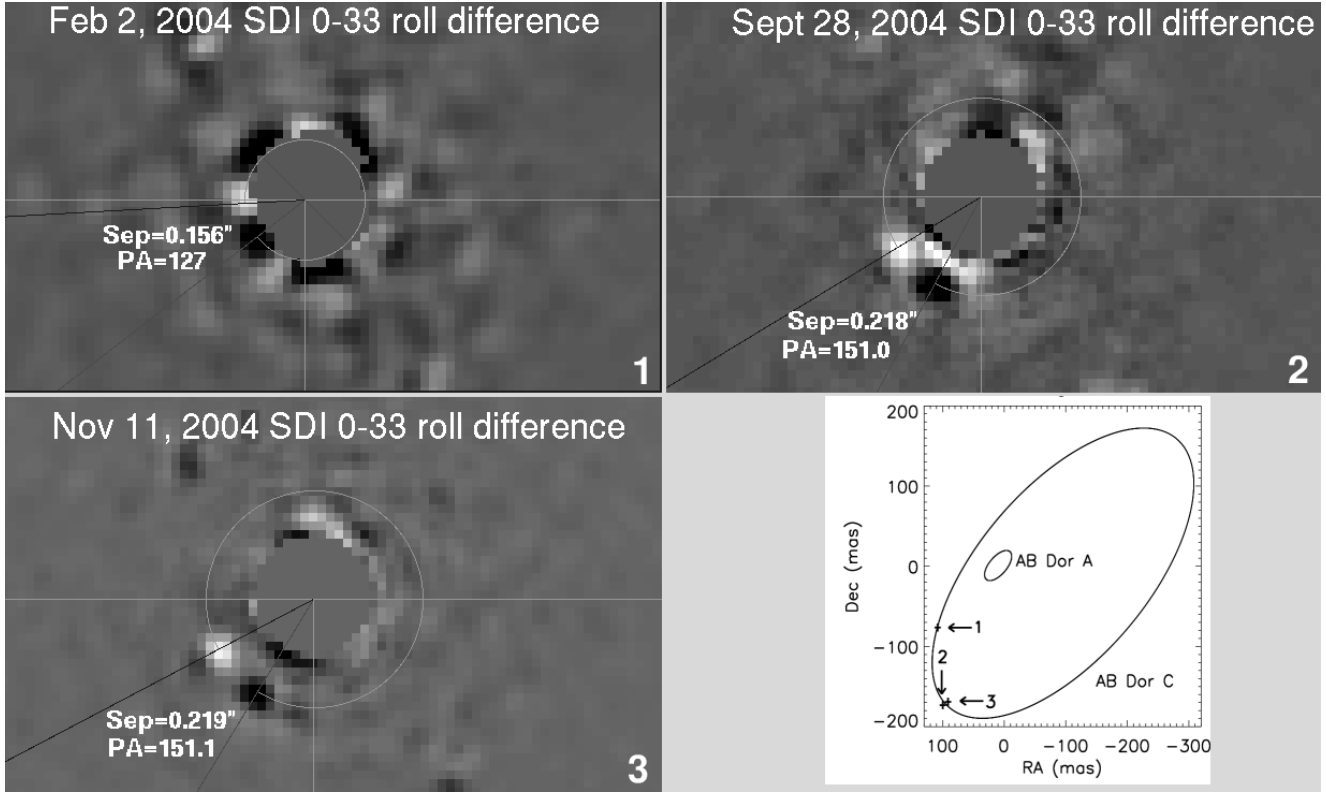}}
\caption{The different epochs of AB Dor C measured during 2004 with the VLT 
SDI device.  Each of the numbered panels shows an individual epoch of SDI 
observations, with images at position angles of 0$^{\circ}$ and 33$^{\circ}$ 
subtracted from each other, showing a positive and negative signal from 
AB Dor C.  The inner pixels of AB Dor A have been intentionally 
saturated, and have been removed from the image.  
The orbital motion of the companion can clearly be seen over 
this span of time.  The bottom-right panel shows these locations against a 
plot of the full orbit of AB Dor C.  The 11 additional VLBI/Hipparchos 
measurements of the reflex motion of AB Dor A (which went into finding the 
orbital solution) are not shown on this plot (See Close et al. 2005).}

\label{orbit01}
\end{figure*}

Our earlier paper (Close et al. 2005) was based on observations 
conducted at the VLT in February 
of 2004.  Since this work was published, we have reduced additional 
Simultaneous Differential Imaging (SDI, see Lenzen et al. 2004) data 
from September and November of 2004.  While data through the narrow-band 
SDI filters do not provide us with any improved photometric information 
(beyond confirmation that between AB Dor A and C, $\Delta$H = 5.20), 
these images do give us additional astrometric data points, allowing us to 
refine the orbit.  Figure~\ref{orbit01} shows each epoch of measuring the 
position of AB Dor C with respect to A.

\begin{figure}
\resizebox{\hsize}{!}
{\includegraphics[width=3in]{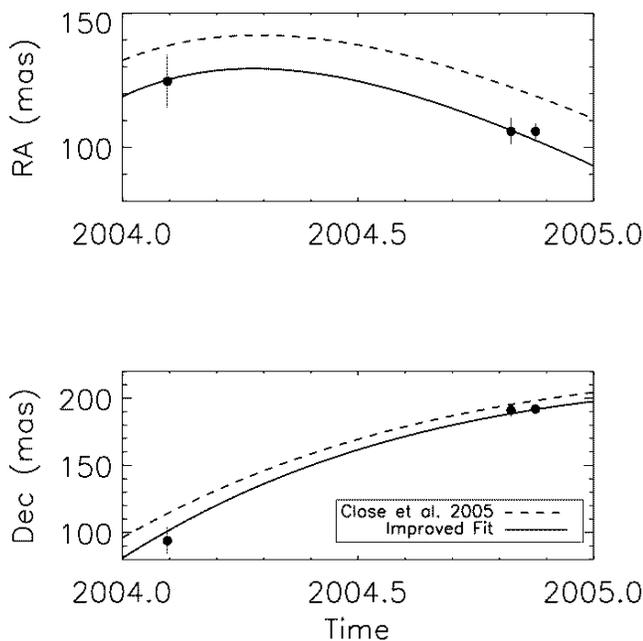}}
\caption{
Detail of the reflex orbit of AB Dor A for the time period of our SDI 
observations.  The previous orbit (Close et al. 2005) is also shown.}
\label{orbit02}
\end{figure}

We fit our three VLT SDI data points for the relative position between AB 
Dor A and C along with the existing VLBI/Hipparcos astrometry for AB Dor A 
(Guirado et al. 1997) to obtain an improved orbit of the reflex motion of AB 
Dor A.  For this fit, we followed the procedure described in Guirado et al. 
2005 (submitted).  The new orbital elements are shown in 
Table~\ref{orbit_tab}, and are mostly similar to those published in 
Close et al. 2005.  The two orbits are compared with respect to the three 
2004 epochs of SDI observations in Figure~\ref{orbit02}.

\begin{table}[ht]
\caption{Our improved parameters for the reflex motion of AB Dor A.}
\label{orbit_tab}
\begin{tabular}{cccc}\hline
Parameter & Value & Error & Units \\
\hline
Period & 11.74 & 0.07 & years \\
Semi-Major Axis & 0.0319 & 0.0008 & `` \\
Semi-Major Axis & 0.476 & 0.012 & AU \\
Eccentricity & 0.61 & 0.03 & \\
Inclination & 66 & 2 & deg. \\
Argument of Periastron & 110 & 3 & deg. \\
Position Angle of Node & 133 & 2 & deg. \\
Epoch of Periastron Passage & 1991.92 & 0.03 & years \\
Mass of AB Dor C & 0.090 & 0.003 & M$_{\sun}$\\
\hline
\end{tabular}
\end{table}

The most immediate consequence of our new orbital fit is that 
the mass of AB Dor C remains at 0.090 M$_{\sun}$.  As reported in 
Guirado et al. 2005, we notice that the error bars shrink from 0.005 
M$_{\sun}$ to 0.003 M$_{\sun}$.  This confidence with which we know 
the mass of AB Dor C makes it an ideal object for calibrating theoretical 
evolutionary tracks.

\section{Discussion}

\subsection{Spectral Type}

\begin{figure}
\resizebox{\hsize}{!}
{\includegraphics[width=3in]{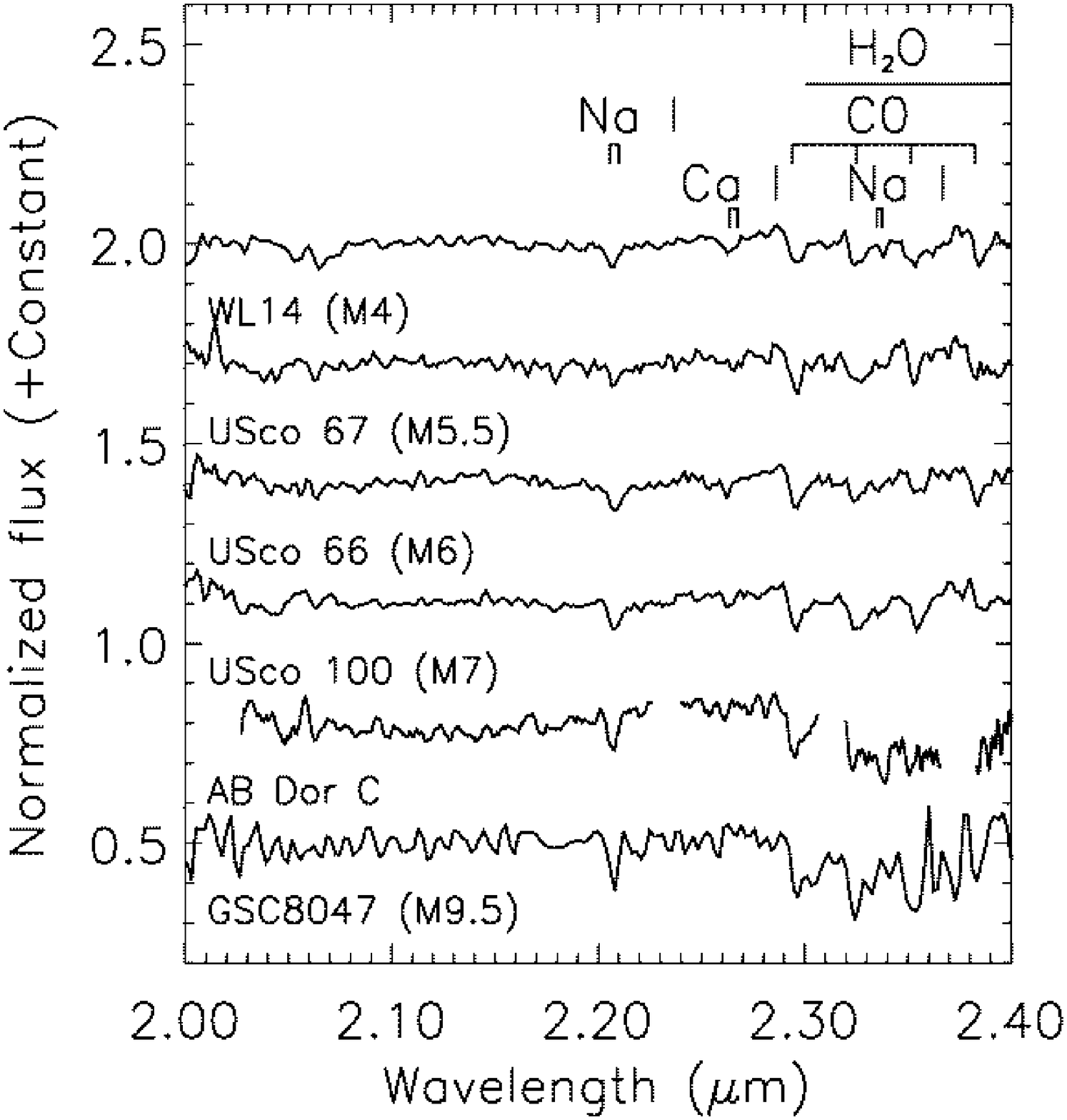}}
\caption{The spectrum of AB Dor C shown against a number of young ($\sim$10 
Myr), 
low-surface gravity objects.  The features seem to be bound between the 
M7 and M9.5 templates.}
\label{lowsg_fig}
\end{figure}

Using our new spectrum, we attempt to refine the determination of the 
spectral type of 
AB Dor C.  Rather than use field objects as our standards (as we did in 
Close et al. 2005), we choose young objects (with lower surface 
gravities) to constrain the spectral type.  Figure~\ref{lowsg_fig} shows 
our AB Dor C spectrum plotted against a variety of young, late-M spectra 
(WL 14, USco 67, USco 66, and USco 100 from Gorlova et al. 
2003, GSC 8047-0232 from Chauvin et al. 2005).  Again, since AB Dor C lacks 
a continuum, we have removed the continuum of all the objects (using the 
same order of polynomial fit) for comparison purposes.  Trends across the 
sequence are clearly visible: as we move to later spectral types, the 
strengths of the Na line, CO breaks, and H$_2$O absorption increases, 
while the Ca line 
weakens.  For all these features, AB Dor C seems best bound between the 
M9.5 and M7 templates at 1$\sigma$, agreeing with the J-Ks $\sim$ 1.3 $\pm$ 
0.4 magnitude color reported in Close et al. 2005.

\begin{figure}
\resizebox{\hsize}{!}
{\includegraphics[width=3in]{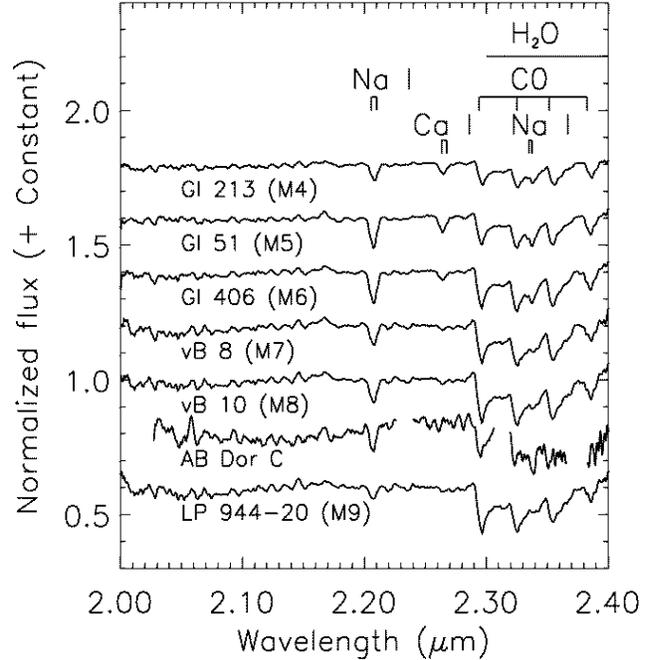}}
\caption{The spectrum of AB Dor C, this time plotted against field 
M dwarfs ($\sim$ 5 Gyr), with higher surface gravities.  Again, a 
spectral type of M8$\pm$1 (1$\sigma$) seems most consistent with 
our spectrum.}
\label{hisg_fig}
\end{figure}

All of the templates used in Figure~\ref{lowsg_fig} are significantly 
younger than AB Dor C.  To properly bound the spectral type, we consider 
field objects, as we did in Close et al. 2005.  Figure~\ref{hisg_fig} 
again shows AB Dor C, now with templates of higher surface gravity (spectra 
from Cushing, Rayner, and Vacca 2005; since these spectra were taken at 
higher resolution, we have smoothed them to match the resolution of AB Dor 
C).  The 
spectra no longer fit as nicely (especially the shapes of the CO features), 
but as before, the spectrum seems to fit best in the sequence between an 
M7 and an M9.

\subsection{The Age of AB Dor}

In Close et al. 2005 it was argued that due to the excess luminosity of AB 
Dor A and C compared to the Pleiades, and A's large Li equivalent width and 
very fast rotation, that the age of the system was 30-100 Myr.  An age of 
50 (-20, +50) Myr was adopted, which was consistent with the 50 Myr 
published age of the AB Dor moving group (Zuckerman, Song, and Bessell 
2004).  Recently Luhman et al. 2005 has suggested a slightly older age of 
70-150 Myr.  However, as Luhman et al. 2005 notes, the AB Dor moving group is 
systematically over-luminous compared to the Pleiades (age 100-120 Myr) by 
$\sim$0.1 magnitudes in M$_{Ks}$ vs. V-Ks plots (see their Figure 1).  

\begin{figure}
\resizebox{\hsize}{!}
{\includegraphics[width=3in]{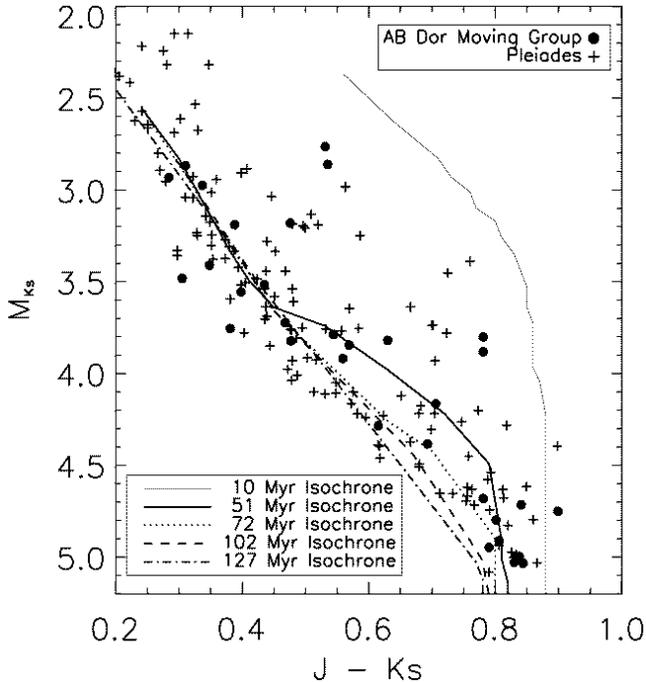}}
\caption{A NIR color-magnitude diagram of medium-mass members of the 
Pleiades (compiled from the literature) and the AB Dor moving group 
(Zuckerman et al. 2004), with the theoretical isochrones of Baraffe et al. 
1998.  The offset between the single star locus of the two groups 
redward of J-Ks $\sim$0.4 suggests a younger age for the AB Dor moving 
group, closer to 70 Myr.}
\label{age_fig}
\end{figure}

We have found similar results with a near-infrared color magnitude diagram, 
as seen in Figure~\ref{age_fig}.  While the two groups of stars appear 
similar for the early-type members (where the isochrones overlap), beyond 
a color of J-Ks $\sim$0.4, the lower main sequence of the AB Dor Moving group 
appears to be above that of the Pleiades by about 0.15 magnitudes.  We 
have run a series of simulations that suggest that only $\sim$10\% of the 
time would a group of Pleiades aged stars appear 0.15 magnitudes above the 
single-star locus as is observed for all the AB Dor group members.

We note that a 0.15 magnitude offset from the Pleiades single star locus 
suggests a group 
age of $\sim$70 Myr from the Lyon group's models (Baraffe et al. 1998).  
We adopt an average age 
of 70 $\pm$ 30 Myr with 1$\sigma$ error bars.  Hence, there is a $\sim$10\% 
chance that the AB Dor moving group is as old as the Pleiades based on 
these color-magnitude diagrams.

Luhman et al. 2005 concludes that their increase in the age of the system 
(from 50 to 120 Myr) implies that the luminosity 
is correctly predicted by the 
models.  But, as we will see in Section~\ref{hrsec}, even if the luminosity 
is close to the predicted value at 
an age of 100 Myr and 0.09 M$_{\sun}$, there is 
still a very large error in the temperature.  Hence, the models will 
overestimate the temperature (or underestimate the mass) of young, low-mass 
objects in the HR diagram regardless of the 70 or 120 Myr age of AB Dor.

\subsection{HR Diagram and Evolutionary Models}\label{hrsec}

\begin{figure}
\resizebox{\hsize}{!}
{\includegraphics[width=3in]{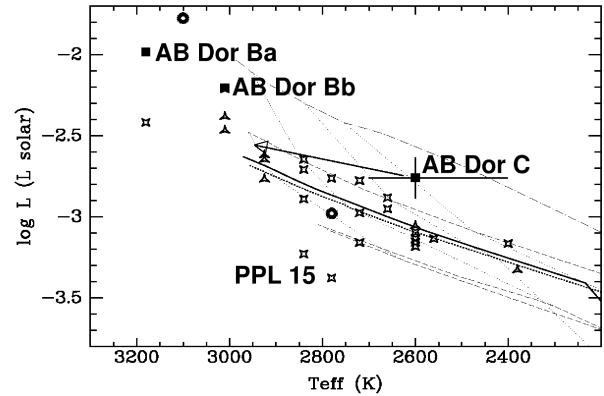}}
\caption{HR diagram showing low-mass Pleiades objects from Martin et al. 
2000 (open stars), other low-mass members of the Pleiades taken from the 
literature (open triangles), and AB Dor Ba/Bb (filled boxes).  Both AB Dor 
Ba/Bb and PPL 15 A/B are shown both as individual objects and as a single, 
blended source (rings).  The dotted vertical lines are 
iso-mass contours for the DUSTY models (from left to right, 0.09, 0.07,0.05, 
and 0.04 M$_{\sun}$), while the more horizontal, dashed lines are the DUSTY 
isochrones (top to bottom, 10, 50, 100, 120, 500, 1000 Myr).  Note that the 
DUSTY models predict a 70-100 Myr object of 0.09 M$_{\sun}$ should be 
$\sim$400 K hotter than observed.  \textit{From the location of AB Dor C 
on the HR diagram, one would derive a mass of 0.04 M$_{\sun}$, a 
factor of 2 underestimate in mass.}  As the temperature and luminosities 
of the Pleiades 
objects in this plot were determined in the same manner used for AB Dor C, 
and these Pleiades points mostly fall along the 
appropriate 120 Myr DUSTY isochrone, 
we are assured that our temperature scale and bolometric correction 
are reasonable.  With 1$\sigma$ error bars, there is a $\sim$99\% 
chance that the DUSTY models underestimate the mass of AB Dor C from the HR 
diagram.}
\label{HRfig1}
\end{figure}

In order to further compare our observations of AB Dor C with the theoretical 
models, we consider an HR diagram with our measured values and the DUSTY 
models.  Using our spectral type of M8 and the absolute Ks from Close et al. 
2005, we can 
derive an effective temperature and bolometric luminosity 
for AB Dor C.  We plot 
AB Dor C in such an HR diagram in Figure~\ref{HRfig1}, along with the 
DUSTY tracks, AB Dor Ba/Bb, and low-mass members of the Pleiades.  We 
compile Pleiades members from Martin et al. 2000, as well as from other 
sources in the literature (Cluster identifications and spectral types 
from Briggs and Pye 
2004, Pinfield et al. 2003, Terndrup et al. 1999, Festin 1998, Martin, 
Rebolo, and Zapatero-Osorio 1996, and Ks-band fluxes from the 2MASS 
catalog)  The bolometric 
luminosities and temperatures for all these objects (AB Dor Ba/Bb, C, and 
the Pleiades members) are derived using Allen et al. 2003 
and Luhman 1999 (dwarf scale), respectively.

As is seen in Figure~\ref{HRfig1}, AB Dor C is overluminous, above the 
Pleiades sequence (as expected from a younger object, $\sim$70 Myr).  
We also show an arrow to its position in the HR diagram predicted 
by the DUSTY models appropriate to its 
age and mass.  

It has been suggested that this overluminosity could be explained if AB Dor C 
were a close, unresolved binary (Martin private communication; Marois et 
al. 2005).  Were this the case, AB Dor C would split into two points in 
Figure~\ref{HRfig1}, and move downward (as AB Dor Ba/Bb and PPL 15 do when 
deblended), appearing consistent with the 
Pleiades locus.  While this interpretation cannot be currently ruled 
out (though we estimate the likelihood to be $<$5\%), we 
will address this issue in more depth in Close et al. 2005, in prep.

\begin{figure}
\resizebox{\hsize}{!}
{\includegraphics[width=3in]{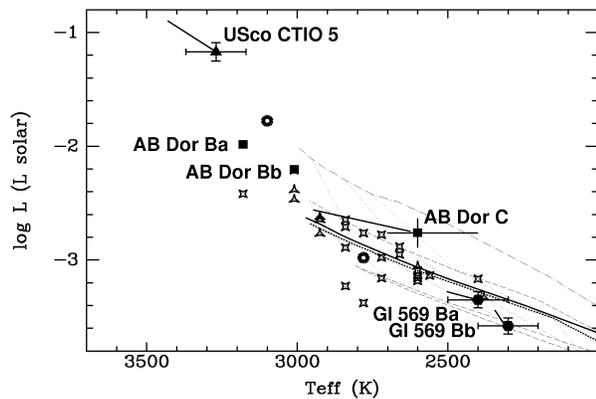}}
\caption{Again, an HR diagram with the Pleiades and certain other young, 
low-mass objects, spanning a range of dynamical masses.  The points with 
error bars mark objects with dynamical masses, with the diagonal lines 
representing the displacement from the measured luminosity and temperature to 
the values predicted by the DUSTY and Next-Gen models.  Upper Sco CTIO 
5 (Reiners et al. 2005), AB Dor C, and Gl 569 Ba/Bb (Zapatero Osorio et al. 
2004) all show a systematic trend where the measured 
HR diagram location is cooler and fainter than the models' predictions.  Seen 
another way, the masses predicted by the models are underestimates of the 
actual masses.  In the 
case of the older (300 Myr) Gl 569 B system, the offset is within the 
1$\sigma$ uncertainties.}
\label{HRfig}
\end{figure}

Finally, we note an overall trend for young, low-mass objects where 
dynamic masses have been measured, as shown in Figure~\ref{HRfig}.  There is 
a global offset to higher luminosities and temperatures for AB Dor C, 
USco CTIO 5, and Gl 569 Ba/Bb.  These results suggest that further work must 
be done to bring theoretical evolutionary tracks in line with observations.

\acknowledgements

We thank Gael Chauvin for providing an electronic copy of the spectrum of 
GSC 8047-0232, and Nadja Gorlova for providing spectra of many young, 
low-mass objects.  We also thank the organizers of the ULMF conference for 
the chance to present this work.  

This publication makes use of data 
products from the Two micron All Sky Survey, which is a joint project of 
the University of Massachusetts and the Infrared Processing and Analysis 
Center/California Institute of Technology, funded by the National 
Aeronautics and Space Administration and the National Science Foundation.

\end{document}